\documentstyle[emulateapj]{article}

\begin{document}
 
\title{A Distant Stellar Companion in the $\upsilon$ Andromedae System}
\author{Patrick J. Lowrance\altaffilmark{1,2}, J. Davy Kirkpatrick\altaffilmark{2} 
\& Charles A. Beichman\altaffilmark{1}}
\altaffiltext{1}{Jet Propulsion Laboratory, 4800 Oak Grove Dr, Pasadena, CA 91109}
\altaffiltext{2}{Infrared Processing and Analysis Center, MS 100-22, California
Institute of Technology, Pasadena, CA 91125}


\begin{center}
{\it Submitted 12 Apr 2002; Accepted 2 May 2002;}
\end{center}

\begin{abstract}

Upsilon Andromedae is an F8V star known to have an extrasolar system of at least 3 
planets in orbit around it. Here we report the discovery of a low-mass stellar 
companion to this system. The companion shares common proper motion, lies at a 
projected separation of $\sim$750 AU, and has a spectral type of M4.5V. The effect 
of this star on the radial velocity of the brighter primary is negligible, but 
this system provides an interesting testbed for stellar planetary formation 
theory and understanding dynamical stability since it is the first multiple 
planetary system known in a multiple stellar system.

\end{abstract}

Subject headings: Astrometry --- binaries: visual --- planetary systems --- 
stars: individual ($\upsilon$ Andromedae) --- stars: low-mass, brown dwarfs

\section{Introduction}

Over the last several years, extrasolar planets have been discovered through
precision Doppler velocity surveys (c.f. Marcy \& Butler 2000). 
Butler et al (1999) reported the first 
evidence for a multiple planetary system around the F8V star $\upsilon$ Andromedae. 
Though the combined M sin$i$ values for these three planets implied a mass 
at least 5 times more than the combined mass of the planets in our own solar 
system, this discovery was heralded as the first evidence 
that our own system of planets is not unique. Currently, there are seven 
systems of multiple planets known, and all are around apparent 
single stars like our own Sun.

During the course of our standardized search for wide companions to stars 
within 25 pc of the Sun, we discovered 
a stellar companion to $\upsilon$ Andromedae 
($=$HD9826:Gl 61; 01:36:47.98 $+$41:24:23.0,V$=$4.10, d$=$13.5pc) at 
an apparent projected separation of $\sim$750AU. In this paper we
present astrometry and spectroscopy to confirm its companionship and 
discuss the possible influence this secondary might have on the planetary system.

\section{Observations}

\subsection{Discovery}

Finding companions at wide separations ($>$ few hundred AU) 
involves searching large pieces of the sky around the nearest stars. 
The Two Micron All Sky Survey (2MASS; Skrutskie et 
al.\ 1997) is currently presenting a wealth of new material with which 
previously unknown low-luminosity companions can be found at wide separations. 
We are in the middle of a 2MASS-based survey around all stars
within the Third Catalogue of Nearby Stars (CNS3; Gliese \& Jahreiss 1991) 
for possible companions out to separations of 0.1 pc. 
The CNS3 includes some
3800 stars believed to lie within 25 pc of the Sun. Companions that are 
M, L and T dwarfs will 
have red optical-to-near-IR colors (R-J). Optical photometry from the 
USNO-A catalog, which is paired up with 2MASS data during routine
2MASS data processing, can be used for an appropriate color constraint. 
Our candidates are 2MASS detections within a search radius equivalent to a physical 
separation of 0.1 pc around the nearby star 
that either have no optical counterparts or have very red $R-J$ colors. Confirmation of 
companionship and classification are then found by astrometric and spectroscopic 
follow-up, respectively. 
Several stellar and substellar companions to nearby, main sequence stars have been found 
in this manner (Kirkpatrick et al 2001a,2001b; Wilson et al 2001; 
Gizis, Kirkpatrick \& Wilson 2001), and 
the search has been standardized around all nearby stars 
to provide more robust statistics
on the frequency of brown dwarf companions to stars.

During our search, we uncovered a candidate in proximity to the bright, nearby F8V star 
$\upsilon$ And. The candidate companion, 2MASSI J0136504$+$412332, has infrared flux and 
colors (J$=$9.39$\pm$0.03; J$-$K$=$0.88) consistent with a mid M dwarf 
(Kirkpatrick \& McCarthy 1994) at the distance of $\upsilon$ And. 
As shown in Figure 1, it lies 55$\arcsec$ from $\upsilon$ And 
at a position angle of 147$^{\circ}$. 

\subsection{Astrometric Confirmation}

The Hipparcos mission (Perryman et al 1997) measured $\upsilon$ And 
to have the proper motion of $\mu _{\alpha} = $ -172.57$\pm$0.52 and 
$\mu _{\delta} = $ -381.03$\pm$0.45 mas per year. 
From positions measured in the POSS-I and POSS-II digitized images and 2MASS (Table 1), 
we find the candidate companion to have a proper motion over a 44.5 year baseline of 
$\mu _{\alpha} = $ -175$\pm$8 and $\mu _{\delta} = $ -390$\pm$6 mas per year. 
Since the two objects' $\mu _{\alpha}$'s agree to 
within 1 $\sigma$ and the $\mu _{\delta}$'s within 2 $\sigma$, 
we conclude they have the same proper motion within the errors. 
We therefore conclude that this object, hereafter referred to as $\upsilon$ 
And B, is a common proper motion companion to the brighter star, which we will refer to as 
$\upsilon$ And A. The known radial velocity planets $\upsilon$ And b, $\upsilon$ And c, 
and $\upsilon$ And d orbit the brighter star $\upsilon$ And A. 
We note that if planetary and stellar companions are discovered 
in the same system, confusion can be avoided if 
stellar objects are noted with a capital letter, 
and planets noted with a lowercase letter.

\subsection{Spectroscopic Confirmation}

The companion was observed spectroscopically on 26 Jul 2001 UT 
using the Double Spectrograph (Oke \& Gunn 1982) on the 
Hale 200in telescope at Palomar Observatory. The instrumental setup 
used the blue grating of 300 lines/mm and the red grating of 316 lines/mm with 
the 2$\farcs$0 slit to provide a wavelength coverage 
from 5200 to 9100 \AA\ at a resolution of 9 \AA. 
Observations were reduced using standard techniques following 
Kirkpatrick, Henry, \& McCarthy (1991) and flux calibrated using 
the standard star LTT 9491 (Hamuy et al 1994). The spectrum is plotted in Figure 2 and 
compared with spectral standards from  Kirkpatrick et al (1991).
The companion's spectrum is best fit by the M4.5V standard's spectrum. 
Based upon the spectral-type mass relation for late F stars (Cox
2000,p. 382), 
we estimate $\upsilon$ And A has a mass of $\sim$ 1.2 M$_{\odot}$, and using the 
spectral-type mass relationship of Henry \& McCarthy (1993), we estimate a mass of 
0.2 M$_{\odot}$ for $\upsilon$ And B.

\section{Discussion}

The radial velocity monitoring group of Marcy, Butler and colleagues 
excluded binaries only if their separations were less than 
2$\arcsec$ (G. Marcy, 2002 pers comm), so this system would still have been 
monitored had the companion been previously known. 
For $\upsilon$ And A, Butler et al (1999) report a measured rms velocity 
semiamplitude of 15.1 m s$^{-1}$, 
after extracting the three planetary orbits from the measurements. 
We calculate a 0.2 M$_{\odot}$ stellar companion at 750 AU would have a 
period of $\sim$ 20000 yrs and a 
velocity amplitude of K $\sim$ 100 m s$^{-1}$ depending on the eccentricity 
and inclination of the orbit. However, the maximum radial 
velocity {\it change} Butler et al might have measured over their 11
year baseline would be a much smaller value, $\sim$ 0.4 m s$^{-1}$ 
(for i=60 and e=0.5). For a highly eccentric (e=0.9) orbit, it could reach
as large as 0.9 m s$^{-1}$. Even so, this companion object would have not been 
detectable in the radial velocity measurements.

Since the announcement of the multiple planets around $\upsilon$ And A, there have been 
many theoretical models of the evolution and stability of three planets 
around a single star. 
While the inner planet is a 'typical' close, hot Jupiter with a small eccentricity (0.04), 
the eccentricities of the outer two increase to 0.18 and 0.41, respectively. 
Interaction between the two outer planets has been proposed as the major influence of orbital 
evolution, but there has been much debate over the stability of such a system. Rivera \& Lissauer (2000), 
who, like Laughlin \& Adams (1999), ignore the innermost planet in the numerical simulations,
found co-planar systems can be stable on small timescales, 
but the cause of the high eccentricities does not seem easily explained by only 
planetary interactions (Jiang \& Ip 2001).  

What influence would a distant stellar companion have on the orbits of a planetary system? 
In the numerical models of the stabilty of 
planets within binary systems, Holman \& Weigert (1999) found certain critical semimajor axes 
for which disruption of the orbit could occur, depending on the eccentricity of the binary stars. 
For the $\upsilon$ And system, the critical orbital radii range from approximately 40AU to 240AU for 
e=0.8-0.0, respectively. While the known radial velocity planets are all well within this limit, 
we note that a circumprimary disk or remnant Kuiper-belt 
could be disrupted by this companion, depending on its inclination and eccentricity. 
In the HD 141569 disk, it has been suggested that the outer gap observed in the debris disk could 
have its origin in gravitational perturbations caused by distant companions (Weinberger et al 2000). 

Another radial velocity planet with a highly eccentric orbit (e$=$0.63) is 16 Cygni Bb. 
It has been suggested that its high eccentricity might be due to interaction with the 
stellar companion, 16 Cygni A, at a projected separation of 800 AU (Mazeh, Krymolowski \& Rosenfeld 1997; 
Holman, Touma, \& Tremaine 1997). The tidal force of the secondary upon the smaller body was 
proposed to induce a modulation of the eccentricity 
over a long time scale ($\sim$10$^8$yr). This would cause a planetary orbit that forms with a small 
eccentricity to slowly increase its eccentricity over the current age of the star ($\sim$ 3 Gyr). 
They calculate the total 
increase is only weakly sensitive to the semimajor axis ratio if the period of the modulation 
is long, and therefore can be a large effect over the lifetime of that system. 

The projected separation of the 16 Cygni binary is consistent with the projected separation 
we find here (750 AU), though $\upsilon$ And B 
is approximately 0.4 the mass of 16 Cygni A. However, the presence of multiple planets in the 
$\upsilon$ And system rules out the kind of effect noted above since the apsidal precession rate
of the planets due to mutual interactions greatly outweighs any induced by the stellar companion 
(Chiang \& Murray 2002). Just as the Moon, because of its proximity, has a larger effect 
on the terrestrial tides than the more massive but more distant Sun, 
the mutual interaction of the planets $\upsilon$ And c 
and $\upsilon$ And d is much larger than any tidal forces from distant $\upsilon$ And B.

\section{Conclusion}

We present astrometric and spectroscopic evidence that 2MASSI J0136504+412332 is 
a M4.5V common proper motion companion to the star $\upsilon$ And, which harbors a multiple 
planetary system. This represents the first system in which multiple planets 
exist within a multiple stellar system, but such a scenario might not be rare. 
There are residual velocity trends hinting of the presence of more than 
one planet around the binary stars 55 Cnc and $\tau$ Boo (Fischer et al 2001). 
There are currently only single planets known around 
the multiple stars 16 Cyg B, 94 Cet, HD 142, HD 179811B, HD 80606, HD 195019, Gl 86, and HD 89774. 

Unifying all aspects of extra-solar planetary formation and evolution 
into a consistent model that accounts for the distribution of known orbital 
parameters remains a daunting task. 
Previously unknown 
stellar or substellar companions might be gravitationally affecting the entire system. 
The complete picture surely includes several factors including disk viscosities, planetary 
interactions, and companions.

The authors wish to thank E. Chiang and the anonymous referee for useful comments which 
helped to clarify the discussion. 
P.J.L. acknowledges support from a National Research Council Fellowship. 
P.J.L., J. D. K., and C. A. B. acknowledge the support of
the Jet Propulsion Laboratory, California Institute of Technology, which is operated under
contract with the National Aeronautics and Space Administration. This publication makes use
of data from the Two Micron All-Sky Survey, which is a joint project of the University of
Massachusetts and the Infrared Processing and Analysis Center, funded by the National
Aeronautics and Space Administration and the National Science Foundation. This research
has made use of the SIMBAD database, operated at CDS, Strasbourg, France.

\begin{deluxetable}{cccc}
\tablecaption{Measured Positions of $\upsilon$ And B} 
\tablenum{1}
\tablehead{\colhead{Epoch (UT)}  &  \colhead{R.A. J2000} & \colhead{Decl J2000} & \colhead{Survey}}
\startdata
16 Sep 1953 & 01:36:51.1   & +41:23:49.9 & POSS I (O) \\
14 Nov 1995 & 01:36:50.5   & +41:23:33.1 & POSS II (N) \\
11 Feb 1998 & 01:36:50.42   & +41:23:32.57 & 2MASS (J) \\
\enddata
\end{deluxetable}

\plotone{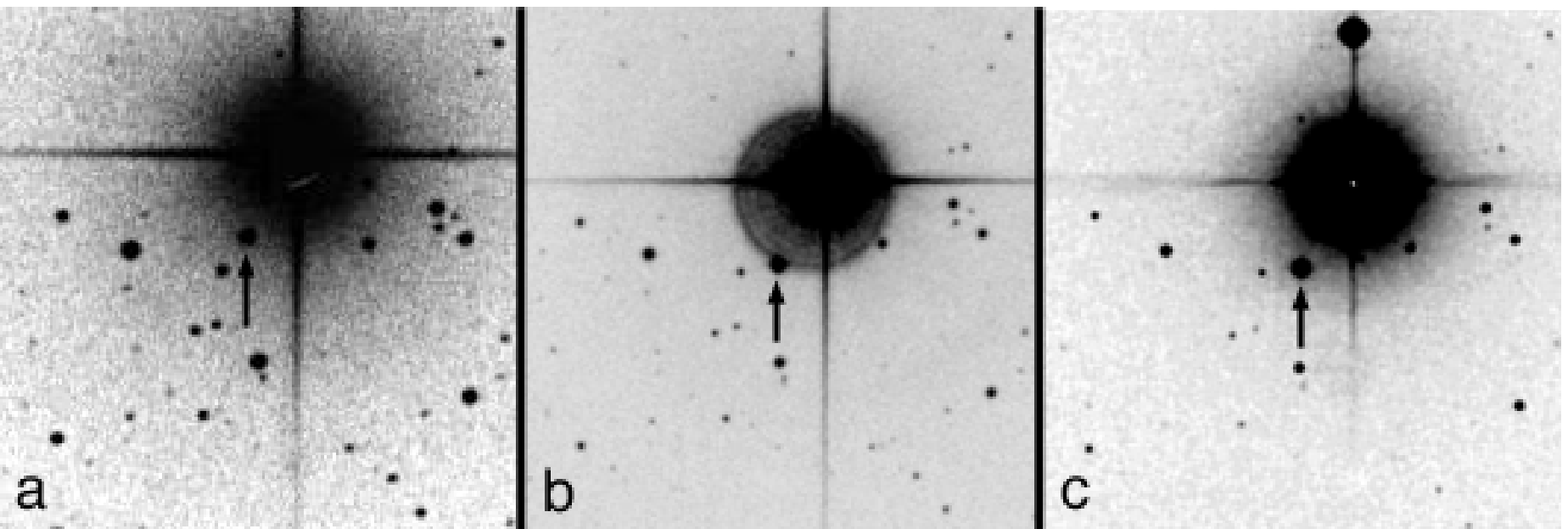}
\figcaption{Discovery images of the $\upsilon$ And B representing the 
(a) POSS-I (O), (b) POSS-II (N), and (c) 2MASS (J) images. As the brighter $\upsilon$ And A 
moves to the southwest over the 44.5 year baseline, so does the fainter companion. (In the 2MASS 
image, the bright object due north of $\upsilon$ And A is a latent image caused by the 2MASS scanning 
procedure, and the glint due south of $\upsilon$ And A and west of $\upsilon$ And B is an internal 
reflection artifact. See http://www.ipac.caltech.edu/2mass/releases/second/doc/explsup.html) }


\plotone{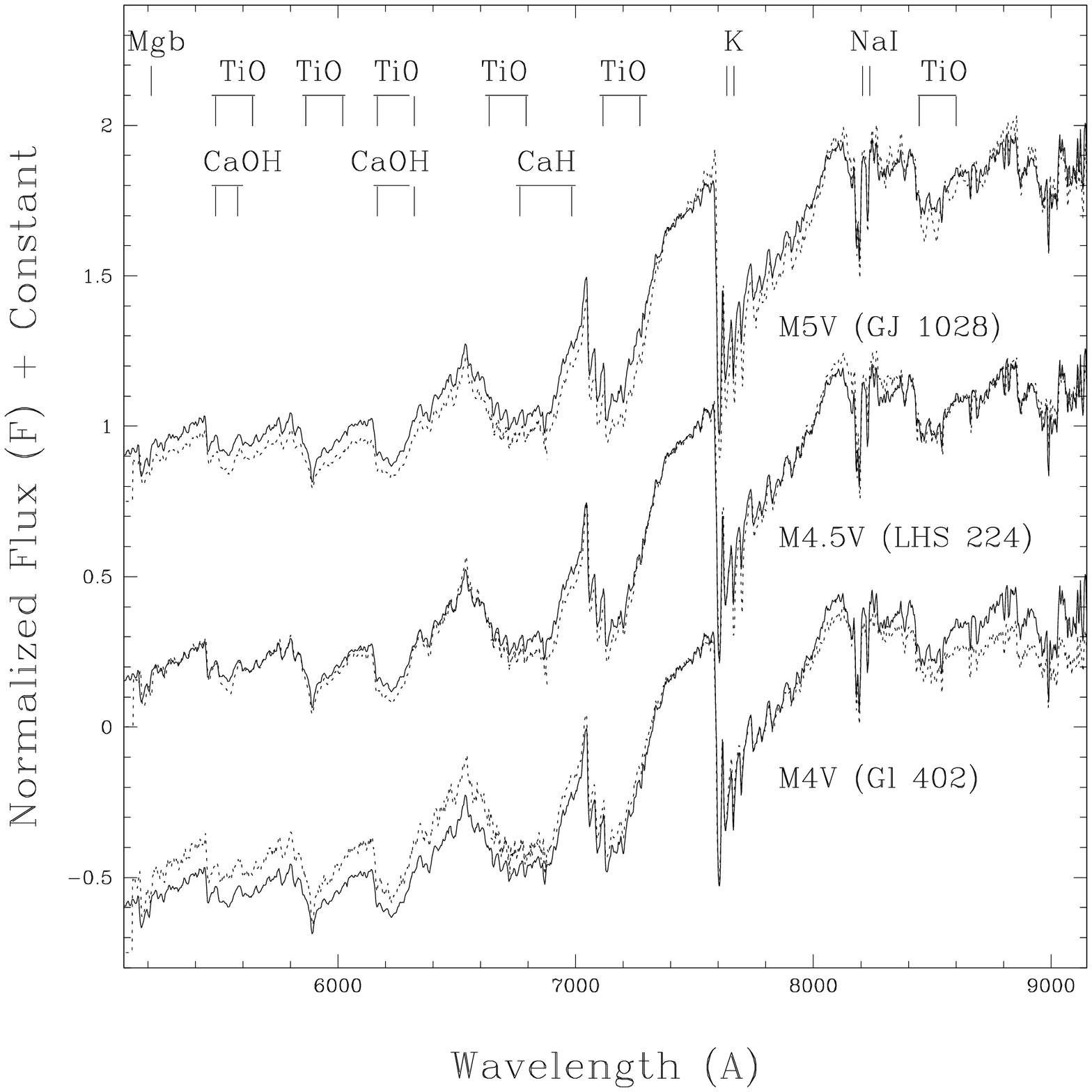}
\figcaption{ Spectrum of $\upsilon$ And B (solid) (in
ergs/s/cm$^2$/\AA)
compared with standard late-type M dwarf spectra (dashed)
(Kirkpatrick et al 1991). The spectra are normalized to the flux at 7500\AA\ and offset by integers. 
The best fit is consistent with an M4.5V spectral type.}

\end{document}